# Cyclic Voltammetry of Ion-Coupled Electron Transfer Reactions for Diagnosing Energy Storage Materials

Keyvan Malaie

Institute of Biochemistry, University of Greifswald, Felix-Hausdorff-Str. 4, 17487 Greifswald, Germany

E-mail: keyvan.malaie@uni-greifswald.de

## Abstract

The methods of Nicholson and Shain and Randles-Ševcík are the paradigms of voltammetry of redox species. However, as they were originally developed for aqueous redox couples, they cannot be directly applied to solid redox films such as those of battery materials. Herein, for the first time, we present a cyclic voltammetry model based on semi-infinite linear diffusion for ion-coupled electron transfer reactions. The simulated CVs contain parameters such as capacity, ion activity, formal potential of proton-coupled electron transfer, and scan rate that are physically more meaningful than those of the current models in characterizing energy storage materials. We apply this model to the $\varepsilon\text{–}MnO_2$ cathode material as a proof of concept and discuss the significance of the simulation parameters for determining the energetics of the underlying phase transitions and the charge storage mechanisms.

According to the present model, two linear regression lines can be established from relatively simple voltammetry experiments for characterizing energy storage materials: 1) The regression line of the CV mid-peak potential vs. Log of ion activity (or pH), in which the slope and the intercept provide information on the type of charge-carrier ions and their solvation state, respectively. 2) The regression line of the capacity vs. $v^{-1/2}$, where the slope and intercept indicate the contributions of bulk and surface charges in thin redox films, respectively.

## Introduction

The theory of cyclic voltammetry (CV) was originally developed in the 1960s for aqueous redox couples by electrochemists with significant contributions from Nicholson and Shain [1–4]. Due to an increasing interest in energy technologies, researchers have considered extending this technique for application in redox thin layers and films [5–9]. In the meantime, with the emergence of new battery concepts, such as multivalent ion insertion compounds in aqueous solution the reliable determination of the charge carrier, crystal behavior, and the role of various other system parameters has become critically important [10,11].

The electrochemical modeling of battery materials is often based on the data from in situ particle imaging, X-ray spectroscopy, or galvanostatic techniques [12–15]. Notably, in the most recent review reports on (multiscale) modeling of battery materials, CV modeling is not even mentioned [16–19]. Yet, CV modeling, primarily employed in electrocatalysis [20–22], offers the advantage of providing currents and potentials simultaneously containing both kinetic and thermodynamic information. Only recently, it has been used for investigating the kinetic effects of electrode particle geometry in Li-ion batteries [23–25] or coupled chemical reactions in Li-S batteries [26].

In the electrochemical modeling of a redox film, the first step is to determine whether the system is controlled by diffusion, which mainly depends on the film's crystal structure, film thickness, and charge-discharge rate. When the system is not under diffusion control, as in the case of some thin films, we have introduced a thermodynamic and a heuristic model to simulate their electrochemical insertion reactions [27,28]. Herein, we assume that the system is controlled by semi-infinite linear diffusion (SILD) and develop a simple CV model that can describe redox films under (quasi)reversible conditions. Indeed, this condition can only be realized for solid-solution redox systems. Furthermore, because porosity and particle geometry can have a significant effect on the electrochemical reaction kinetics [29,30] we will not use our model for analyzing the reaction kinetics but the thermodynamic and charge storage properties. Thus, the assumption of SILD is justified here. In addition, we develop the model around the proton-coupled electron transfer (PCET), but it is equally valid for other ions.

## Results and Discussion

**Diffusion Model.** Scheme 1 depicts the three-phase system of the redox film. We consider a fast PCET at the reaction plane at $y$ = 0 as described by eq. 1. The symbols $b$ and $n_e$ are the stoichiometry coefficients of the proton and electron.

$$R \rightarrow O + bH^+ + n_e e^- \quad (1)$$

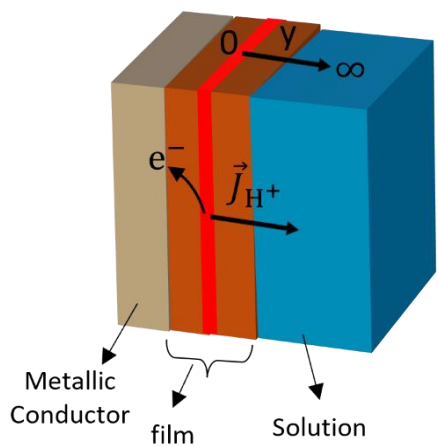

**Scheme 1** A macroscopic view of the flux of H⁺ ions ($\vec{J}_{H^+}$) within the redox film under SILD condition during the electrochemical oxidation of R to O at $y$ = 0, i.e., the reaction plane.

The diffusion of protons within metal oxides ($D$) is typically several orders of magnitude lower than that in an aqueous solution [31–33]. Thus, it should determine the overall proton transport. In addition, since the diffusion of the protons from/to the redox centers is coupled with the evolution of the redox species [34], the mass transport can be expressed in redox species concentration according to Fick's second law. Furthermore, the total mass of the redox species is conserved during the reaction, i.e., $C_O + C_R = C^*$. Hence, we can express Fick's second law in terms of the mole fraction of O (designated with $x$) in the one-dimensional space of $y$ as follows (Sec. 2, ESI):

$$\frac{\partial x(y,t)}{\partial t} = D\frac{\partial x(y,t)}{\partial y^2} \quad (2)$$

To solve this partial differential equation (PDE), three boundary conditions are required, as defined below.

**Boundary Conditions.** The first two conditions arise from the assumption of SILD with only R initially being present. Thus, at time zero the amount of O is zero, and at a distance sufficiently far from the reaction plane the amount of O is zero, as well, as expressed by eq. 3 and 4, respectively.

$$\lim_{t \to 0} x(y,t) = 0 \quad (3)$$

$$\lim_{y \to \infty} x(y,t) = 0 \quad (4)$$

Besides, at the reaction plane, i.e., $y = 0$, if the electrochemical reaction is fast, it follows the following Nernst equation (See Sec. 3, ESI):

$$E_{eq}(0,t) = E_t^{0'} + \frac{RT}{n_e F} \ln \frac{x(0,t)}{1 - x(0,t)} - \frac{2.302 bRT}{n_e F} \text{pH} \quad (5)$$

The formal potential $E_t^{0'}$ depends on the reorganization energy for the coordination shell of the metal center during electron transfer to the metallic conductor and the energy of proton transfer between the film and the solution. $b/n_e$ denotes the ion-to-electron transfer. Other symbols have their normal definitions (Sec. 1, ESI). Eq. 5 can be rearranged as follows:

$$x(0,t) = \frac{1}{1 + e^{-fE(0,t)}} \quad (6)$$

In which $f$ denotes the $n_e F/RT$ term. $E(0,t)$ is a function of the linear potential program of the CV (Sec. 3, ESI). Eq. 6 is an expression of the Fermi-Dirac statistics for O and R species under equilibrium [35] that is obtained simply by rearranging the Nernst equation. This equation constitutes the last boundary condition required for solving the PDE in eq. 2.

**Solving the Diffusion Problem.** A step-by-step analytical and numerical solution of eq. 2 under the above boundary conditions is presented in Sec. 4 of ESI which follows the numerical method of Nicholson and Shain [1]. The result is presented in eq. 7:

$$g(1)\sqrt{n} + \sum_{i=1}^{n-1} \sqrt{n-i}[g(i+1) - g(i)] = \frac{\sqrt{\pi D}}{2\sqrt{\delta}} \frac{1}{1 + e^{-fE(\delta n)}} \quad (7)$$

In eq. 7, time is discretized as $t = \delta \times n$ with $\delta$ and $n$ denoting the length and number of steps, respectively. $g(n)$ is an unknown function that should be numerically evaluated for each $n$. Finally, the current can be expressed based on $g(t)$ as follows:

$$i(t) = \frac{\partial x(0,t)}{\partial t} q_t = g(t) q_t \qquad (8)$$

Here, $q_t$ denotes the capacity (C/cm²) of the redox film.

Fig. 1a exhibits the linear potential program, $E(t)$, of the CV for pH=0. Fig. 1b displays the $i(t)$ function computed based on eq. 7 and 8 for $q_t$ = 1 C/cm², and Fig. 1c shows a plot of the numerical values of $i(t)$ vs. $E(t)$. This CV model verifies the two familiar features of a reversible electrochemical reaction. It has a diffusion tail following the CV peaks, which is a result of the SILD assumption, and it displays a Peak-to-peak separation of ~57 mV for a one-electron transfer reaction. Below, we discuss the properties of the as-developed CV model and its application in characterizing energy storage materials.

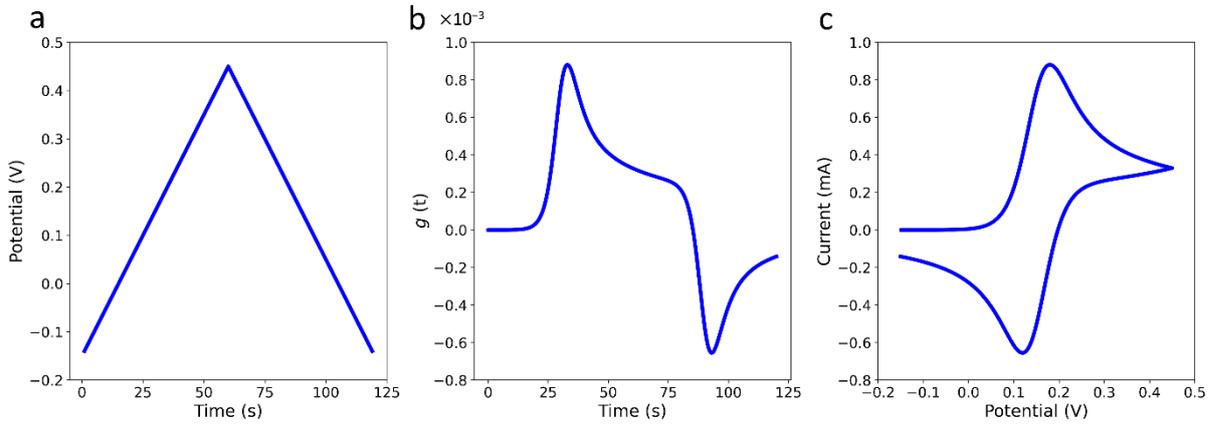

**Fig. 1** Voltammetric model of a redox film limited by diffusion. **a)** the linear potential program of $E(t)$ **b)** the numerical presentation of the $g(t)$ function vs. time (eq. 7), and **c)** the plot of $g(t)*q_t$ vs. $E(t)$. (For the values of model parameters see Tables S1 and S2)

**Diffusion Coefficient**: Fig. 2a presents the effect of different solid-state diffusion coefficients on the CV models, respectively, verifying the linear dependence of peak current ($i_p$) on $D^{1/2}$. It should be noted that the solid-state diffusion of ions may depend on the state of charge ($x$) and hence may not be constant [36,37]. In addition, it is assumed in this model that the oxidation and reduction diffusion coefficients are equal. This is a reasonable assumption because first the same ion is inserted or deinserted, and second, the dependence of the equilibrium potential on unequal diffusion coefficients is very weak following the relationship $E \propto 0.059 \log(D_R/D_O)^{1/2}$. For instance, for the familiar $Fe^{2+}/Fe^{3+}$ couple in aqueous solution, the $D_R/D_O$ ratio is only about 1.1 translating to a deviation of 2.8 mV from the potentials considered in the present model [38,39].

**pH and $b/n_e$.** Another prediction of the model is that the CV curves shift in the negative direction by $b/n_e \times 0.059$ V for each unit increase in pH. Fig. 2b presents the simulated CV curves for three different pH values for a one-electron transfer reaction, exhibiting a slope of 59 mV/dec. This method can be used to determine the underlying reaction for redox films in aqueous solutions.

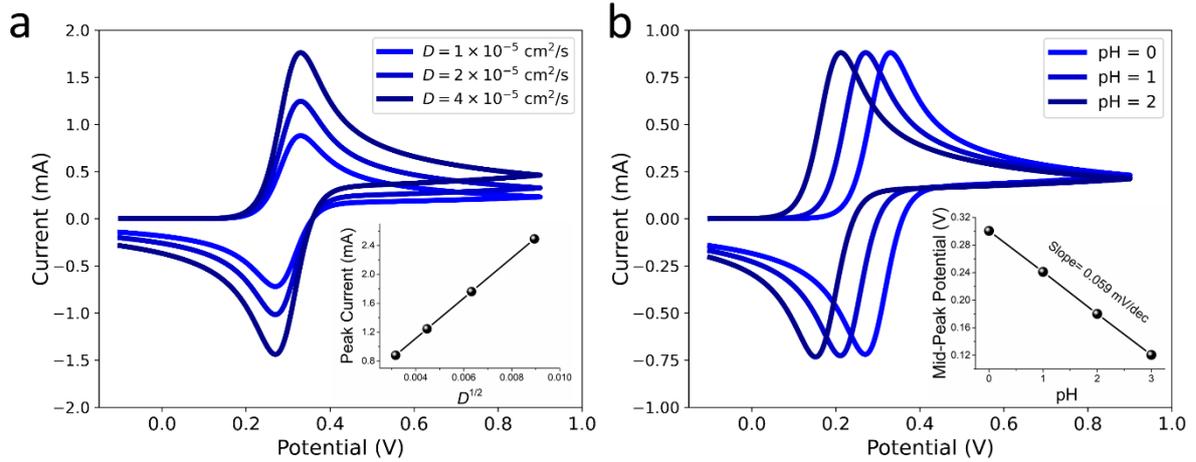

**Fig. 2** The effect of different solid-state diffusion coefficients **(a)** and solutions with different pH **(b)** on the CV models generated based on the eq. 7 and 8. for a redox film. (For the values of model parameters see Tables S3 and S4)

Fig. 3a shows the experimental CV curves of an ε-MnO$_2$ coating in aqueous solutions with different pH. The CV curves display the expected features of a (quasi)reversible system. They show a diffusion tail, a linear $i_p/v^{1/2}$ relationship, and peak-to-peak separations that are slightly larger than expected by 20–70 mV due to ohmic polarization [40]. The medium of the two oxidation and redox peaks is called the mid-peak potential $E_{mp}$ and corresponds to $x = 0.5$ in eq. 5. For the quasi-reversible systems as in the case of the present ε-MnO$_2$ electrode, the exact determination of the $E_{mp}$ may require an extrapolation method for $E_{mp}$ points determined at different scan rates [40]. Fig. 3b exhibits the $E_{mp}$ values estimated for the ε-MnO$_2$ electrode at different pH values. Such diagrams are also known as Pourbaix diagrams. The $E_{mp}$ - pH regression lines have slopes of 62 and 123 mV/dec for alkaline and acidic solutions, respectively, corresponding to $b/n_e$ values of 1 and 2 at 25 °C (see eq. 5). Thus, according to eq. 1, they reveal the following underlying reactions, respectively:

$$\text{MnOOH} \rightleftharpoons \text{MnO}_2 + \text{H}^+ + e^- \quad (9)$$

$$\text{Mn}^{2+} + 2\text{H}_2\text{O} \rightleftharpoons \text{MnO}_2 + 4\text{H}^+ + 2e^- \quad (10)$$

eq. 9 is the main cathode reaction in alkaline batteries [41], and both reactions are involved in the cathodes of Zn–ion batteries [42–44]

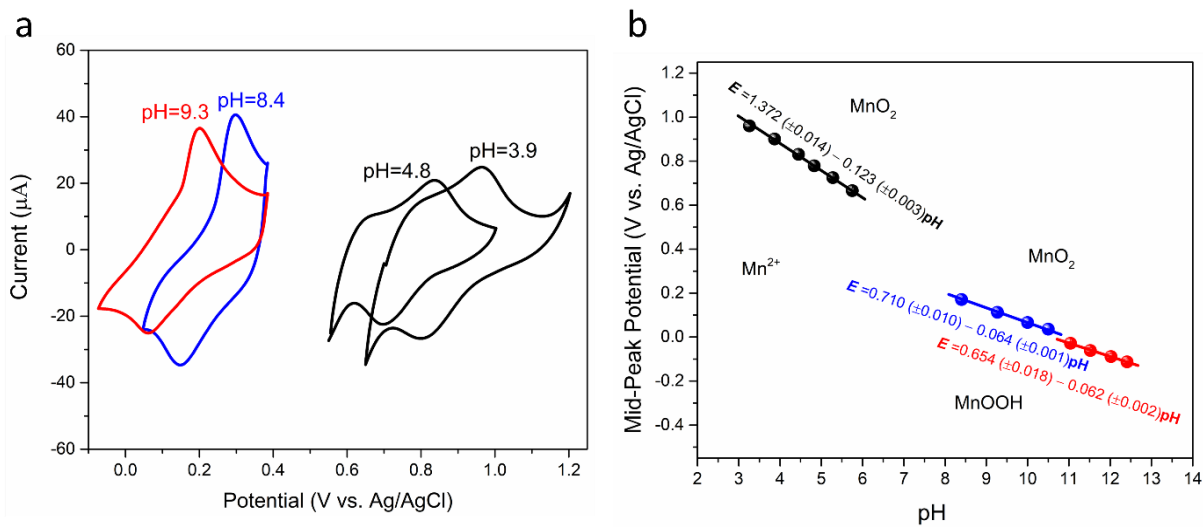

Fig. 3 a) The effect of solution pH on the CV curves of ε-MnO$_2$ at 2.5 mV/s. The acidic solution is 0.02 M acetate buffer containing 0.01 mM Mn$^{2+}$ and the alkaline solution is 0.1 M ammonium buffer. b) The mid-peak potentials of the CV curves plotted against solution pH. (Adapted from Malaie et al.[40] licensed under CC BY 4.0)

**Formal Potential ($E_t^{0'}$):** The $E_{mp}$ value at pH=0 is the formal potential of the proton-coupled electron transfer for the redox couple (see eq. 5). It is simply the intercept in the plots of Fig. 3b converted to the standard hydrogen electrode (SHE) scale. For the ion-insertion electrodes, the $n_e F E_t^{0'}$ term is the sum of the molar free energies of electron transfer (ET) and ion transfer (IT). This energy is also known as Bond Dissociation Free energy (BDFE). For the MnO$_2$/Mn$^{2+}$ couple, the formal potential contains the formation energy of ε–MnO$_2$, which has been determined to be −427.3 kJ/mol [40]. This energy reveals that ε–MnO$_2$ is less stable than the most thermodynamically stable phase (β-MnO$_2$) by about 35 kJ/mol. The thermodynamic instability of ε–MnO$_2$ is expected because as a highly-defective crystal almost 50% of the metal sites are vacant [45]. Similarly, the formal potentials of insertion compounds can be used to determine their BDFE [46] and possibly even the individual contributions of ET and IT energies [40].

Additionally, by using the as-determined BDFEs in a thermochemical cycle, it is possible to discern the solvation state of the insertion ion, i.e., if the ion carries solvent with itself into the crystal. Scheme 2 depicts a thermochemical cycle of the three-phase insertion electrode shown previously in Scheme 1. Since BDFE is equal to the sum of the free energies of ET and IT, the thermochemical cycle can be completed with the work function (W) of the metallic conductor, the solvation energy of H$^+$, and the lattice energy of the MnO$_2$. Using this method, the solvation energy of H$^+$ was determined to be ~1014 kJ/mol comparable with the accepted value of ~1056 kJ/mol [40]. The lower solvation energy measured for H$^+$ can indicate an incomplete desolvation of the H$^+$ when inserting the crystal. The incomplete desolvation of the insertion ion has been observed for other compounds in aqueous solutions, as well [47].

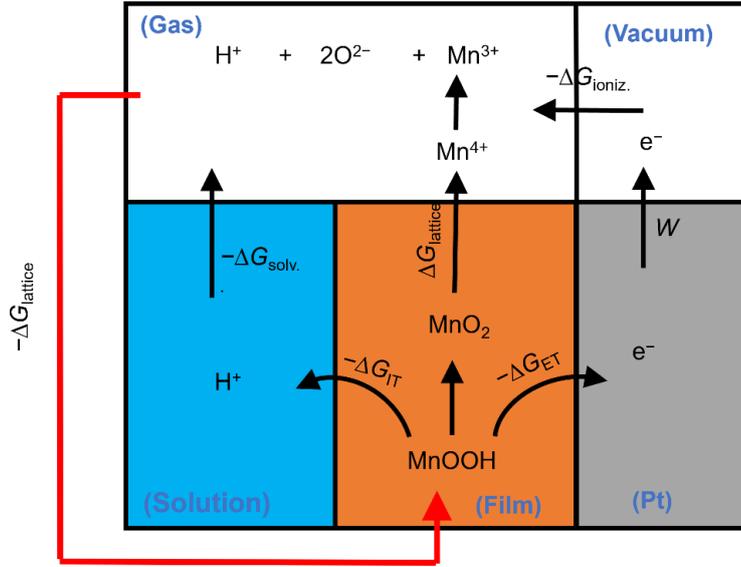

**Scheme 2** A thermochemical cycle for the electrochemical proton (de)insertion in the ε-MnO₂ film in conjunction with an aqueous solution. (*W*: work function, Δ$G_{ET}$ and Δ$G_{IT}$ denote standard free energies of electron transfer and ion transfer)

**Scan rate (*v*)**: Contrary to the other parameters of the present model which are thermodynamic properties of the redox system, the *v* is a method parameter. The effect of *v* on the voltammetric signals of redox films is complicated because increasing *v* can reduce electrode capacity mainly due to ohmic polarization (energy dissipation) and even change the reaction pathway in two-phase materials [28,48]. Fig. 4a shows the effect of *v* on the CV models and their peak currents (inset) when the capacity remains constant. These $i-v^{1/2}$ relationships (also called *b*-value analysis) have been used widely and often subjectively to distinguish diffusion-limited and capacitive processes [49]. Nonetheless, we suggest a $q_t-v^{-1/2}$ analysis as described next.

**Capacity (*q*ₜ) and Film Thickness (*L*):** A last but not least important parameter in the model is capacity. It is inherently linked to the apparent film thickness, i.e., the distance traveled by the ions via diffusion during the measurement time (*t*). This relationship can be expressed according to eq. 11 [50]:

$$\frac{q_t}{A} = \frac{q_s}{A} + q_{bulk}L \qquad (11)$$

Where $q_s$ and $q_{bulk}$ are the charges stored at the surface and in the bulk of the electrode, respectively. *L* is the apparent thickness being equal to $\sqrt{Dt}$. In a voltammetry measurement, *t* can be expressed based on *v* according to *t*=*RT*/*vF*. Thus, from eq. 11:

$$\frac{q_t}{A}(v) = \frac{q_s}{A} + q_{bulk}\sqrt{\frac{DRT}{vF}} \qquad (12)$$

Eq. 12 is a simple proof of the empirical Trasatti analysis in which the slope of the $q_t-v^{-1/2}$ line is an unknown constant [51,52]. Fig. 4b shows a plot of eq. 12 for a redox film with $D = 1 \times 10^{-5}$ at *v* values from 160 mV/s to 10 mV/s. The apparent thickness evaluated from eq. 11 is indicated on the plot, as well. Thus, the slope and the intercept of the $q_t-v^{-1/2}$ line provide the bulk capacity (limited by diffusion) and the surface capacity of the film, respectively. This analysis is particularly important for nanostructured coatings [53,54], and it can be implemented in two ways: 1) if the films are sufficiently thin, i.e., usually below 50 nm, the apparent and the real thicknesses are the same. Then, by measuring the real thickness, a linear regression of $q_t$–$L$ (eq. 11) provides the surface and bulk charges as recently performed for $TiO_2$ thin films by Xiao et al. [50]. 2) if the solid-state diffusion coefficient is known or determinable from another technique [55], eq. 12 can be used accordingly. Experimentally, at very high scan rates or for thick films, the $q_t$–$L$, $q_t$–$v^{-1/2}$, as well as the $i_p$–$v^{1/2}$ plots become nonlinear and show a trade-off in capacity (See Fig. S1, ref. [56] and [57]). This effect is due to the ohmic polarization through the film including the contacts and the solution, which is not considered in the present diffusion model.

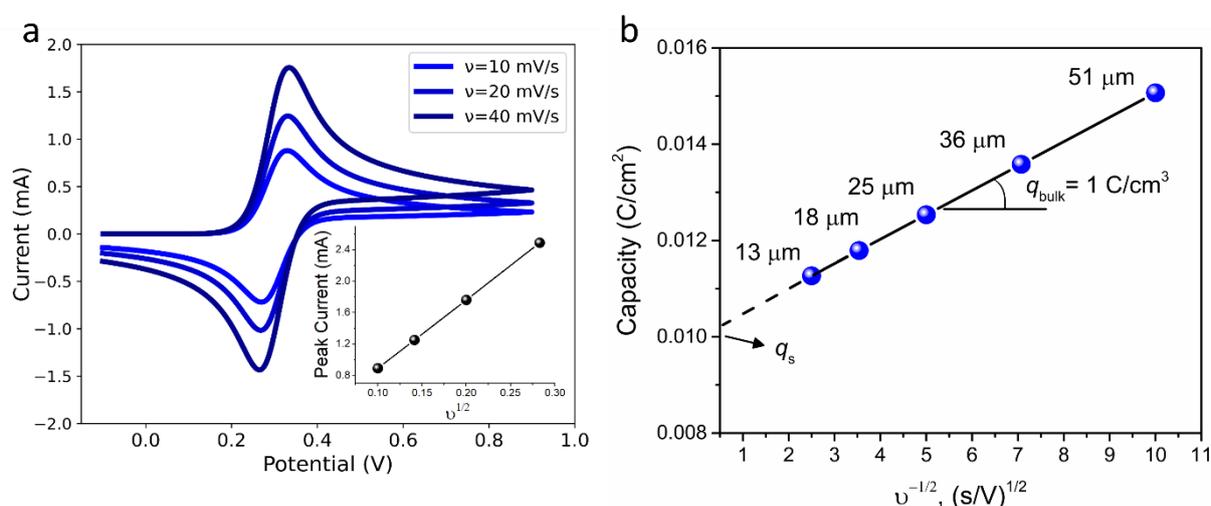

**Fig. 4 a)** The effect of different scan rates on the CV and the peak current (inset) according to eq. 7 and 8. **b)** the effect of electrode thickness and scan rate on the electrode capacity (eq. 11 and 12). ( for the values of model parameters see Table S5) .

Finally, table 1 presents a comparison of the present model with that of Nicholson and Shain. First, it shows that the present model is greatly simplified in methodology because it requires less number of boundary conditions and parameters. Second, the parameters are physically more meaningful for analyzing energy storage materials.

**Table 1** Comparison of the parameters in the present model with those of Nicholson and Shain

| Parameters in the Nicholson & Shain Model | Equivalent Parameters in the Present Model |
|---|---|
| Concentrations of soluble redox species of O and R | Mole fraction of O |
| Two diffusion coefficients | One solid-state diffusion coefficient |
| Electron number | Proton-to-electron number ($b/n_e$) |
| Standard potential of electron transfer | Formal potential of proton-coupled electron transfer |
| Bulk concentration | Electrode Capacity |
| - | Activity of the insertion ion in the liquid phase |

## Conclusion

A voltammetric model based on semi-infinite linear diffusion is introduced for characterizing battery (and pseudocapacitive) materials. The model parameters include solid-state diffusion coefficient, pH, proton-to-electron number, ion-coupled electron transfer formal potential, scan rate, and capacity. They are described briefly, applied to the ε-$MnO_2$ as an example, and finally compared with those of existing voltammetry models.

Most significantly, according to the present model, two linear regression lines can be established from rather simple experiments as follows: 1) The regression line of the mid-peak potential vs. Log of ion activity (or pH): the slope and the intercept of this line inform on the type of charge-carrier ions and their solvation state (or crystal formation energy), respectively. 2) The regression line of the capacity vs. $v^{-1/2}$: the slope and intercept of this line can be used to determine the contribution of bulk and surface charge storage in thin redox films.

Finally, the model applies to solid-solution redox films in which the phase properties of the film do not change substantially during the charge or discharge. The next step in this work is to consider other electrode/particle geometries and consider ohmic effects in the model, which necessitates the application of a Deep Neural Network (DNN) method over the present numerical method to solve the diffusion problem.

## Conflicts of Interests

There are no conflicts to declare.

## Acknowledgment

KM acknowledges the Alexander von Humboldt Foundation and the University of Greifswald.

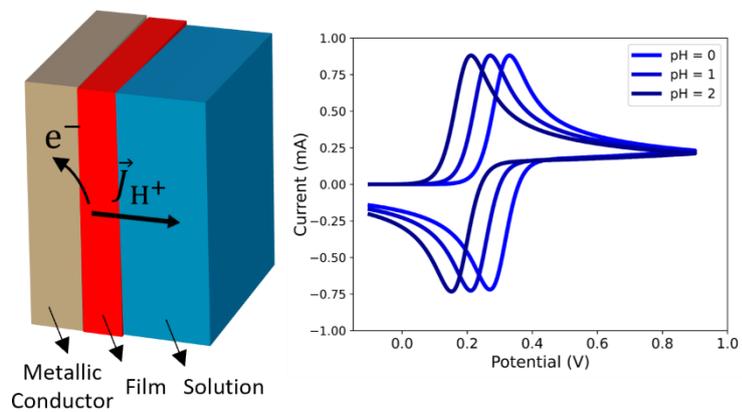

For Table of Contents Only

Supporting Information

# Cyclic Voltammetry of Ion-Coupled Electron Transfer Reactions for Diagnosing Energy Storage Materials

Keyvan Malaie

Institute of Biochemistry, University of Greifswald, Felix-Hausdorff-Str. 4, 17487 Greifswald, Germany

E-mail: keyvan.malaie@uni-greifswald.de

## 1. Experimental Methods

All potentials mentioned in the theory section are referenced against a suitable reference electrode with a negligible junction potential. All the experiments were carried out at 298 K, and this temperature was assumed in all the CV simulations. The values of the Faraday constant and the gas Constant are assumed 96485 C/mol and 8.314 J/(mol*K), respectively.

All data analysis, visualization, and evaluations were performed using Spyder, an open-source Python environment provided by Anaconda, Inc. The numerical computation of the partial differential equations (PDEs) was also carried out in Spyder.

The detailed experimental conditions for the ε-$MnO_2$ and $Ni_3S_2$ electrodes can be found in Refs. [1] and [2]. The details of the process for calculating the free energy terms of the thermochemical cycle in Scheme 2 can be found in the supporting information of Ref. 1.

## 2. Diffusion Problem based on Mole Fraction

We assume that diffusion is the only mechanism of mass transport within the solid redox phase. In addition, the double-layer charging and ohmic polarization are not considered. Furthermore, the model assumes that the oxidative and reductive ionic diffusion coefficients are equal (i.e., $D_O = D_R$) and constant during the charge and discharge. It follows from this assumption that the local concentrations of the ions at any point are conserved according to $c_O + c_R = c^*$ or $x_O + x_R = 1$, in which $c$ and $x$ denote species concentration and mole fraction, respectively. Hence, mass transport within the solid phase can be expressed based on $x$ (mole fraction of the O species as described by eq. 1) as follows:

$$\frac{\partial c(y,t)}{\partial t} = D\frac{\partial c(y,t)}{\partial y^2} \quad \text{(S1)}$$

$$\frac{\partial x(y,t)}{\partial t} = D\frac{\partial x(y,t)}{\partial y^2} \quad \text{(S2)}$$

$D$ denotes the solid-state diffusion coefficient of the O species, and $y$ is the length dimension in which diffusion takes place.

## 3. Derivation of the Third Boundary Condition

The proton transfer at the redox film-solution interface under equilibrium conditions follows a partition reaction defined with a partition coefficient $K_d$ according to the eq. S3 and S4:

$$\text{H}^+(\text{aq}) \rightleftarrows \text{H}^+(\text{f}) \quad \text{(S3)}$$

$$K_d = \frac{a_{\text{H}_f^+}}{a_{\text{H}_{aq}^+}} \quad \text{(S4)}$$

$a$ denotes the activities of H$^+$ in the film and solution. Under constant activity coefficients, we have:

$$K_d' = \frac{c_{\text{H}_f^+}}{c_{\text{H}_{aq}^+}} \quad \text{(S5)}$$

where $K_d'$ is the formal partition coefficient.

At the reaction plane, i.e., at $y = 0$, for a fast electron transfer, the reaction in eq. 1 follows the Nernst equation. By using formal potential ($E^{0'}$) to account for activity coefficients as follows, it follows:

$$E_{eq} = E^{0'} + \frac{RT}{n_e F}\ln\frac{c_o c_{\text{H}_s^+}^b}{c_R} \quad \text{(S6)}$$

$E_{eq}$ is the equilibrium potential of the electrode vs. some reference electrode, and $E^{0'}$ is the formal potential of the electron transfer for the O/R couple. Using mass conservation, eq. S6 can be expressed in terms of the mole fraction:

$$E_{eq} = E^{0'} + \frac{RT}{n_e F}\ln\frac{x}{1-x} + \frac{bRT}{n_e F}\ln c_{\text{H}_f^+} \quad \text{(S7)}$$

Where $x$ is the mole fraction of the O at the reaction plane at $y = 0$. The $H_f^+$ species do not enter the mass balance equations (similar to acid-base equilibria). Next, $c_{H_f^+}$ in eq. S7 can be substituted from eq. S5 as follows:

$$E_{eq} = E_t^{0'} + \frac{RT}{n_e F} \ln \frac{x}{1-x} - \frac{2.302 bRT}{n_e F} pH \tag{S8}$$

In which $E_t^{0'}$ includes the p$K_d'$ and the activity coefficients, in addition to the standard potential of electron transfer for the O/R couple.

Eq. S8 can be rearranged as follows:

$$x(0,t) = \frac{1}{1 + e^{-fE(t)}} \tag{S9}$$

In which $f$ is equal to $RT/n_e F$ and $E(t)$ in CV scans has the following expressions:

$$E(t) = E_i - E_t^{0'} + \frac{2.302 bRT}{n_e F} pH + vt, \qquad 0 < t \le \lambda \tag{S10}$$

$$E(t) = E_i + 2v\lambda - E_t^{0'} + \frac{b2.302 RT}{n_e F} pH - vt, \qquad \lambda \le t \tag{S11}$$

Where $v$ and $\lambda$ denote the potential scan rate and the time when the potential scan is reversed, respectively. Eq. S9, S10, and S11 constitute the third boundary condition.

## 4. Solving the Diffusion Problem

Eq. S2 can be converted to a homogeneous ordinary differential equation (ODE) by performing the Laplace transform and applying the first boundary condition (eq. 3) as shown in eq. S12:

$$\frac{\partial \bar{x}(y,s)}{\partial y^2} - \frac{s}{D} \frac{\partial \bar{x}(y,s)}{\partial y} = 0 \tag{S12}$$

Eq. S12 has the following general solution:

$$\bar{x}(y,s) = A(s)e^{-\sqrt{s/D}y} + B(s)e^{\sqrt{s/D}y} \tag{S13}$$

$A(s)$ and $B(s)$ are the coefficients of the general solution. Application of the second boundary condition (eq. 4) to eq. S13 results in eq. S14:

$$\bar{x}(y,s) = A(s)e^{-\sqrt{s/D}y} \tag{S14}$$

Next, the convolution theorem is used to perform the inverse Laplace of eq. S14 and obtain *x(y,t)*. However, we are only interested in the mole fractions of O at y = 0 at different times, i.e., *x(0,t)*. Thus:

$$x(0,t) = \frac{1}{\sqrt{\pi D}} \int_0^t \frac{g(\tau)\,d\tau}{\sqrt{t-\tau}} \qquad (S15)$$

The symbol τ is a dummy variable. The integral term in Eq. S15 can be solved either with the semi-integral method of Oldham et al. [3] or with the numerical method of Nicholson and Shain [4]. In the numerical method, eq. S15 can be expressed as a summation of discrete steps by describing the continuous variable τ as the number of steps *n* multiplied by step length *δ* with unit of second, and the integration limits become 0 and *n*. Next, the integration term can be replaced by its summation using integration by parts and then substituting for the Riemann-Stieltjes integral, resulting in eq. S16:

$$x(0,\delta n) = \frac{2\sqrt{\delta}}{\sqrt{\pi D}} \left[ g(1)\sqrt{n} + \sum_{i=1}^{n-1} \sqrt{n-i}[g(i+1) - g(i)] \right] \qquad (S16)$$

Eq. S16 is the solution of the diffusion equation (eq. S2) based on an unknown step function *g(n)*. Finally, by substituting for *x(0,δn)* in eq. S16 from the third boundary condition (eq. S 9, S10, and S11):

$$g(1)\sqrt{n} + \sum_{i=1}^{n-1} \sqrt{n-i}[g(i+1) - g(i)] = \frac{\sqrt{\pi D}}{2\sqrt{\delta}} \frac{1}{1 + e^{-fE(\delta n)}} \qquad (S17)$$

eq. S17 is an implicit function where *g(n)* can be determined for any arbitrary *n*. The *g(n)* function can be expressed as a function of the summation of previous terms and *n* and numerically solved with an iterative code.

## 5. Model Parameters Used in the CV Simulations

Table S1. Values of the linear potential program used in plotting **Fig. 1a**

| Parameter | Value | Unit |
|---|---|---|
| Potential scan rate ($v$) | 0.01 | V/s |
| Initial potential ($E_i$) | 0 | V |
| Formal potential ($E^{0'}$) | 0.15 | V |
| Reversal Time ($\lambda$) | 60 | S |

Table S2. Values of the parameters in the $g(n)$ function used for plotting **Fig. 1b**

| Parameter | Value | Unit |
|---|---|---|
| Diffusion coefficient ($D$) | $1 \times 10^{-5}$ | cm²/s |
| Electrode Area ($A$) | 0.1 | cm² |
| pH | 0 | |
| Time step ($\delta$) | 0.1 | S |
| Number of steps ($n$) | 1200 | - |
| Reversal Time ($\lambda$) | $n \times \delta / 2$ | S |

Table S3. Values of the model parameters used for CV simulations in **Fig. 2a**

| Parameter | Value | Unit |
|---|---|---|
| Potential scan rate ($v$) | 0.01 | V/s |
| Initial potential ($E_i$) | 0 - 0.2 | V |
| Formal potential ($E^{0'}$) | 0.15 | V |
| Diffusion coefficient ($D$) | $1 \times 10^{-5}$<br>$2 \times 10^{-5}$<br>$4 \times 10^{-5}$ | cm²/s |
| Electrode Area ($A$) | 0.1 | cm² |
| Ion-to-electron number ($b/n_e$) | 1 | - |
| pH | 0 | |
| Time step ($\delta$) | 0.1 | S |
| Number of steps ($n$) | 1200 | - |
| Reversal Time ($\lambda$) | $n \times \delta / 2$ | S |

**Table S4.** Values of the model parameters used for CV simulations of **Fig. 2b**

| Parameter | Value | Unit |
|---|---|---|
| Potential scan rate ($v$) | 0.01 | V/s |
| Initial potential ($E_i$) | 0 - 0.2 | V |
| Formal potential ($E^{0'}$) | 0.15 | V |
| Diffusion coefficient ($D$) | $1 \times 10^{-5}$ | cm²/s |
| Electrode Area ($A$) | 0.1 | cm² |
| Ion-to-electron number ($b/n_e$) | 1 | - |
| pH | 0<br>1<br>2 | - |
| Time step ($\delta$) | 0.1 | S |
| Number of steps ($n$) | 1200 | - |
| Reversal Time ($\lambda$) | $n \times \delta / 2$ | S |

**Table S5.** Values of the model parameters used for CV simulations in **Fig. 4a**

| Parameter | Value | Unit |
|---|---|---|
| Potential scan rate ($v$) | 0.01<br>0.02<br>0.04 | V/s |
| Initial potential ($E_i$) | 0.2 | V |
| Formal potential ($E^{0'}$) | 0.3 | V |
| Diffusion coefficient ($D$) | $1 \times 10^{-5}$ | cm²/s |
| Electrode Area ($A$) | 0.1 | |
| Ion-to-electron number ($b/n_e$) | 1 | - |
| pH | 0 | - |
| Time step ($\delta$) | 0.1 | S |
| Number of steps ($n$) | 2000<br>1000<br>500 | - |
| Reversal Time ($\lambda$) | $n \times \delta / 2$ | - |

## 6. Experimental $i_p - v^{1/2}$ Curve

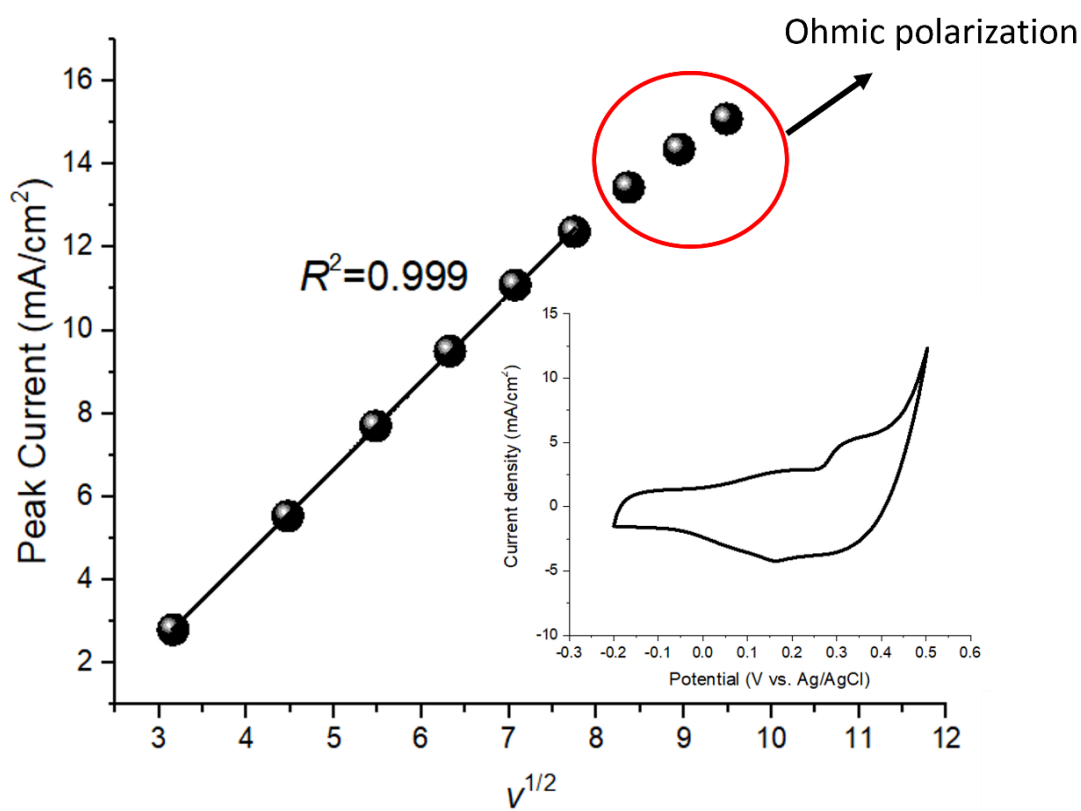

**Fig. S1.** The relationship of CV peak current vs. scan rate for a nanostructured Ni$_3$S$_2$ coating in a 3 M KOH. For experimental details see ref. [2]